\documentstyle[aps,preprint]{revtex}
\begin{document}
\title{Anatomy of nuclear shape transition in the relativistic mean field
theory}
\author{Tapas Sil$^1$, B. K. Agrawal$^2$, J. N. De$^1$ and S. K.
Samaddar$^2$}

\address{$^{(1)}$ Variable Energy Cyclotron Centre,
1/AF, Bidhannagar, Calcutta - 700064, India}

\address{$^{(2)}$ Saha Institute of Nuclear Physics,
1/AF, Bidhannagar, Calcutta - 700064, India}

\date{\today}

\maketitle

\begin{abstract}
A detailed microscopic study of the temperature dependence of the shapes of some
rare-earth nuclei is made in the relativistic mean field theory. 
Analyses of the thermal evolution of the single-particle orbitals and
their occupancies leading to the collapse of the deformation are presented.
The role of the non-linear $\sigma-$field on the shape transition
in different nuclei is also investigated; in its absence the shape
transition is  found to be sharper.

\vskip 0.5cm

PACS number(s) 21.10.Ma, 21.60.Jz, 27.70.+q
\end{abstract}

\newpage
\section{Introduction}

Interaction between nucleons in a nucleus may give rise to preferred
orientations of the single-particle orbitals leading to deformed intrinsic
shapes of the nuclei. With thermal excitations, they undergo a phase
transition to spherical shapes. This has been studied in the
experiments on the shapes of giant dipole resonances (GDR) built on excited
states \cite{sno,gaa}.  The understanding of the mean field shape evolution
with temperature has been attempted in a macroscopic approach
\cite{lev,alh} generally referred to as the Landau theory of phase
transitions. They have also been studied in microscopic framework like
finite-temperature non-relativistic Hartree-Fock \cite{bra,que} and
Hartree-Fock-Bogoliubov (HFB) approaches \cite{goo1,goo2,egi1} with the
pairing-plus-quadrupole (P$+$Q) interaction. The shape transition
temperature is found to be  in the domain of $\sim 1.0 $ to $1.8$ MeV for
the rare-earth nuclei.  Here the model Hamiltonian is simplistic, the model
space is small, an inert core is assumed and moreover, the role of the
Coulomb field is taken into consideration in an effective manner. Recently,
we have studied \cite{agr1,agr2} this same phenomenon in the framework of
relativistic mean field theory (RMF). Here the model space is sufficiently
large, all the nucleons are treated on equal footing and the Coulomb
interaction is properly accounted for. It is then found that the shape
transition temperature is noticeably higher. Except for some calculations
in the s-d shell nuclei \cite{mil1,mil2}, nearly all 
the calculations have been done for nuclei in the rare-earth region; it is
found that the deformation undergoes a sudden collapse at the shape
transition temperature. Grossly, one understands the dissolution of the
deformation with temperature in these nuclei from the following: shell
structure leads to the population of the deformation-driving states, the
so-called intruder states producing the static ground state deformation;
their depopulation with gradual heating restores the spherical symmetry. 

The aim of the present paper is to analyse in more microscopic details the
collapse of the deformation with temperature. In doing so, we also  explore
whether the sudden collapse observed in the rare-earth nuclei is universal
or system-specific. The stability towards the deformed ground state for the
axially symmetric nuclei that we consider is given by the arrangement of
the single-particle orbitals of good projection quantum number; the
evolution of the energies of these single-particle orbitals with
temperature, particularly those near the Fermi surface and also their
occupancy evolution lead to a self-consistent rearrangement of the orbitals
and thus can reveal in detail why the deformed nuclei undergo a shape
transition. 

The non-linear $\sigma-$coupling  term in the effective Lagrangian has been
introduced \cite{bog} to reproduce properly the finite size effect,
particularly the surface properties of nuclei.
Since the spin-orbit coupling term is sensitive to the fall of the density
at the nuclear surface, the structure of the Fermi surface is expected to
depend on the non-linear  term and thus might affect the ground state
deformation and its thermal evolution. These  aspects are also investigated
using a parameter set (HS) where the non-linear $\sigma$-coupling term is
absent. 

A brief discussion of the theoretical framework we employ is given in
section II. The results and discussions are presented in section III. The
concluding remarks are given in section IV. 

\section{Formalism}
The details of the formalism are given in Ref.\cite{agr1}. For the sake of
completeness, however, we write down the effective Lagrangian density 
describing the nucleon-meson many body system followed by a very 
brief discussion. The Lagrangian density is given by
\begin{eqnarray}
\label{lag}
{\cal L}&=& \bar\Psi_i\left ( i\gamma^\mu \partial_\mu - M\right )\Psi_i
+ \frac{1}{2} \partial^\mu\sigma\partial_\mu\sigma - U(\sigma)
- g_\sigma \bar\Psi_i \sigma\Psi_i\nonumber\\
&& - \frac{1}{4}\Omega^{\mu\nu}\Omega_{\mu\nu}
+\frac{1}{2}m_\omega^2\omega^\mu \omega_\mu - g_\omega \bar\Psi_i \gamma^\mu
\omega_\mu\Psi_i 
-\frac{1}{4}\vec{R}^{\mu\nu} \vec{R}_{\mu\nu} + \frac{1}{2} m_\rho^2 \vec{\rho}^\mu\vec{\rho}_\mu\nonumber\\
&&- g_\rho \bar\Psi_i \gamma^\mu\vec{\rho}_\mu\vec{\tau}\Psi_i
-\frac{1}{4}F^{\mu\nu}F_{\mu\nu} - e\bar\Psi_i \gamma^\mu \frac{(1-\tau_3)}{2} A_\mu\Psi_i
.
\label{Lag}
\end{eqnarray}
The mesons included in the description are scalar-isoscalar $\sigma$,
vector-isoscalar $\omega$ and vector-isovector $\rho$ mesons. The arrows in
Eq.(\ref{lag}) indicate isovector quantities. The scalar self-interaction
term $U(\sigma)$ of the $\sigma$ meson is taken to be non-linear
\begin{equation}
U(\sigma) = \frac{1}{2}m_\sigma^2 \sigma^2 + \frac{1}{3}g_2 \sigma^3 + 
\frac{1}{4}g_3\sigma^4
.
\end{equation}
The nucleon mass is $M$; $m_\sigma$, $m_\omega$ and $m_\rho$ are the meson
masses, $g_\sigma$, $g_\omega$ and $g_\rho$ are the coupling constants for
the mesons and $e^2/4\pi=1/137$ is the fine structure constant. 
The field tensors for $\omega$ and $\rho$  are
given by $\Omega^{\mu\nu}$ and $\vec{R}_{\mu\nu}$; for the 
electromagnetic field, it is $F^{\mu\nu}$. Recourse to the variational
principle followed by the mean field approximation treating the field as
$c-$numbers results in the Dirac equation for the nucleon and Klein-Gordon
type equations for the mesons and the photon. For   the static case, along
with the time-reversal invariance and charge conservation, the  equations
get simplified. The resulting equations, known as the relativistic Hartree
equations or RMF equations alongwith the BCS approximation for inclusion of pairing 
are solved to yield the fields and the single-particle energies.

The self-consistent solutions are obtained using the basis expansion
method \cite{gam,rin}; this yields the quadrupole deformation $\beta_2$ and
also the single-particle states as a function of temperature. 

\section {Results and  Discussions}

For the values of the coupling constants and the masses of the mesons and
the nucleons occuring in the Lagrangian density given by Eq. (\ref{lag}), we
choose the NL3 parameter set. This parameter set reproduces the ground
state properties of finite nuclei very well; it yields also the
compressional properties satisfactorily \cite{lal}. The single-particle
states are calculated using spherical oscillator basis with twelve shells.
The values of the chemical potential and the pairing gap at a given
temperature are determined using all the single-particle states upto
$2\hbar\omega_0$ above the Fermi surface without assuming any core. In order
to check the convergence of the calculation, we have enlarged the basis
space from twelve to twenty shells and have extended the model space to
include single-particle states upto $3\hbar\omega_0$ above the Fermi
surface. For this extended model space, the pairing strength  is
adjusted to reproduce the ground state pairing gap. The change in the
values of the observables are found to be insignificant due to this
extension of the basis and the model space even at the highest temperature
we consider. Continuum corrections are neglected as their effects are found
to be negligible \cite{agr2,agr3,bon} in the temperature domain of
relevance (less than $\sim 3.0$ MeV); the shape transition temperatures for
the systems considered are within this range. 

In the rare-earth region, we have studied the thermal evolution of the
shapes of even-even isotopes of $Sm$, $Gd$ and $Dy$; for neutron number 86
and 88, their quadrupole deformation $\beta_2$ as a function of temperature
is presented in Fig. 1. The following aspects  are apparent from
the figure: (i) for the isotopes differing by two neutrons, the difference
in the ground state deformation $\beta_2^0$ is 
significant, (ii) addition of two protons does not
increase  $\beta_2^0$ significantly (compare
$\beta_2^0$ for $^{148}Sm$ and $^{150}Gd$), (iii) there is a close
correlation between $\beta_2^0$ and the  critical tempeature $T_c$
\cite{agr2} and that (iv) $\beta_2$ remains nearly constant with
temperature and then there is a sudden collapse of the deformation near
$T_c$ in a temperature window of $\sim 0.2$ MeV. To elucidate these
phenomena more clearly in  microscopic  terms, we present results for the
thermal evolution of the single-particle levels and their occupancies in
the next couple of figures for $^{148}Sm$ and $^{150}Sm$. These
representative cases  guide one to draw inference about the shape evolution
of the other nuclei.
In Figs. 2-5, the single particle levels near the Fermi
surface for the protons and the neutrons in $^{148}Sm$ and $^{150}Sm$ are
displayed as a function of temperature. The Fermi energies are shown by the
dashed lines; in the range of the temperature concerned, they are nearly
constant. The ground state is axially symmetric and the levels are
degenerate in $\pm K$, $K$ being the projection quantum number. The level
structures remain practically unaltered upto $T\sim 0.75$ MeV for
$^{148}Sm$ and  upto $\sim 1.0$ MeV for $^{150}Sm$ beyond which the different
$K-$levels pertaining to a definite ($nlj$)-orbit start to converge. At the
temperature $T_c$ (1.15 MeV for $^{148}Sm$ and  1.6 MeV for $^{150}Sm$ )
they become degenerate signifying transition to spherical symmetry.
It may be mentioned that in the simplistic (P+Q) model, the phase
transition to a spherical shape occurs relatively earlier; in a recent
HFB calculation in an extended model space with the realistic Gogny
force \cite{egi2}, it has however been  found that the transition 
temperature is in keeping with that found in our calculations in the
RMF framework \cite{agr1}.

Examination  of the single-particle spectra is helpful in understanding why
the two neutron addition to  $^{148}Sm$ increases the ground state
deformation whereas the two proton addition does not alter it
significantly. Because of pairing, the occupancies of the states near the
Fermi surface is partial for a nucleus in its ground state. From Figs.
3 and 5 one sees that the two neutron addition increases
maximally the population of the $^{1/2} f_{7/2}$ state (for the orbitals,
we have used the notation $^Kl_j$);  the populations of
the states $^{1/2}h_{9/2}$ and $^{3/2}h_{9/2}$ are also significantly
enhanced. All these orbitals are intrinsically highly prolate which
explains the increase in the ground state deformation of $^{150}Sm$. 
(Because of
axially symmetry in the even-even nuclei we study, the projection $K$ is a
good quantum number and is degenerate  with ($-K$) because of time
reversal invariance.  The states $^Kl_j$ just mentioned are the ones that
have the largest amplitude for the orbital with projection $K$).
From Figs. 2 and 4 we also note that the 
deep lying $^{7/2}g_{7/2}$ proton orbital in $^{148}Sm$
which is intrinsically oblate moves up closer to the Fermi
surface in $^{150}Sm$ and thus is less populated. This also contributes to
the increase in the prolate deformation for $^{150}Sm$. Examination of the
single-particle spectra for $^{148}Sm$ and $^{150}Gd$ (not shown here)
shows that  the addition of  two protons does not practially affect the
neutron occupancies  near the Fermi surface; For protons only the
$^{3/2}h_{11/2}$ orbitals get  noticeably  more populated thus 
slightly enhancing the deformation in the ground state for the 
$^{150}Gd$ nucleus compared to $^{148}Sm$ as shown in Fig. 1.
The ground state deformation properties of $Dy$  isotopes can be explained
in a similar fashion. 

The suddenness of the transition from a prolate to the spherical shape at
the temperature $T_c$ can be understood as follows: as the temperature
rises, the states above the Fermi surface get increasingly more populated
at the expense of the states below. This causes a self-consistent
reorganization of the single-particle  field because of which the
single-particle states evolve with temperature as depicted in Figs. 2-5.
With increase in temperature for $T$ above $\sim$ 0.5 MeV, one finds
from Fig.6 for the proton orbitals in $^{148}Sm$ a tendency for 
equalisation of the single-particle occupancies for different 
$K-$states emanating from the same $j-$orbital. Around $T_c$, 
one could readily see the sharp drop in the occupancies of the 
prolate orbitals with an accompanied enhancement of the same for the 
oblate orbitals resulting in the rapid collapse of the deformation.
The neutron orbitals (not shown here) have  also the similar
features. All these arguments apply also for the other rare-earth nuclei
considered.

To examine whether the sudden collapse of the deformation as observed in
the rare-earth nuclei is a generic feature or system-specific, one needs to
consider the shape evolution of some other nuclei in a different mass
region. For this purpose, we study two isotopes of  zinc nuclei, namely
$^{64}Zn$ and $^{66}Zn$. As all the features related to shape evolution  in
these isotopes are found to be very similar, we discuss  
only the nucleus $^{64}Zn$. In
Fig. 7, we display the thermal evolution of the deformation
$\beta_2$ for this nucleus. To draw a comparison of the shape evolution
with the rare-earth systems, we also display in the same figure the
temperature dependent deformation for the nucleus $^{150}Sm$. It is
seen that the fall-off of the deformation with temperature of $^{64}Zn$
is somewhat slower. This is magnified in Fig.8 where 
the heat capacity per particle for the
two systems $^{150}Sm$ and $^{64}Zn$ are 
displayed. For  each curve, the twin peaks at
lower temperature correspond to the neutron and proton pairing transitions.
For the ligher nucleus, this transition is relatively more prominent and is
at a little higher temperature as expected. The peaks at  higher 
temperatures ($\sim 1.6$ MeV for $Sm$ and $\sim 2.0$ MeV for $Zn$) 
refer to the shape transition; whereas it is very prominent
for the rare-earth system, it is rather weak and looks more like
a plateau for the $^{64}Zn$ nucleus
signalling a relatively smoother shape transition 
to the spherical configuration. It is further that though the ground
state deformations of these two nuclei are nearly the same,  the
transition temperature for $^{64}Zn$ is significantly higher. 
Comparison of the thermal evolution of the
level structures of $^{64}Zn$ (shown in Fig.9) with those of $^{150}Sm$ 
displayed in Figs.4 and 5 may help in understanding the delayed
shape transition in $^{64}Zn$. For $^{150}Sm$, in the vicinity of 
the transition temperature, the single-particle orbitals of opposite
intrinsic deformations cross each other near the Fermi surface 
hastening ite approach towards a spherical configuration. The higher
$T_c$ in $^{64}Zn$ may be partially attributed to the absence of
these level crossings.

It has been mentioned that the non-linear $\sigma-$coupling
term is necessary to explain the surface properties of finite nuclei
\cite{bog}. Though in the absence of this term, the nuclear matter
binding energy  and the saturation density are reproduced, the binding 
energies of finite nuclei can not be obtained properly \cite{ser}.
The fall-off of the nucleon density  profile at the surface is found to be
stiffer without the non-linear term. Since the spin-orbit splitting  is
proportional to the density gradient, this splitting is then likely to   be
larger without the non-linear $\sigma-$term. This may affect the
single-particle level structure to which the shape evolution is expected to
be sensitive.  In order to study this effect on the ground state
deformation and the thermal shape evolution of nuclei, calculations have
been  performed with a parameter set that does not contain the non-linear
$\sigma-$coupling term. The HS parameter set \cite{hor} is chosen for
this purpose. In Fig. 10, the results for the temperature evolution
of deformation $\beta_2$ are shown for two representative cases, namely,
$^{148}Sm$ and $^{150}Sm$. For comparison, $\beta_2$ with the NL3
parameter set for $^{150}Sm$ is also displayed in this figure.  It is observed
that with the HS set the ground state deformation is larger, the transition
temperature smaller and that the collapse of the deformation is somewhat
faster. The sharper deformation collapse is corroborated in the thermal
evolution of the specific heat with and without the non-linear
$\sigma-$term in the  representative case of $^{150}Sm$ as shown in Fig. 11
(upper panel). This is amplified further in the case of
$^{64}Zn$ as shown in the bottom panel of the figure.
The change in the ground state deformation 
obtained with the HS parameter set can
be understood from the examination of the single-particle spectra near the
Fermi surface. For example, as shown in Fig. 12
for the proton single-particle spectra for $^{150}Sm$, 
the strong prolate deformation-producing
$^{1/2}h_{11/2}$ and $^{3/2}h_{11/2}$ orbitals are significantly deeper
down the Fermi surface as compared to those  displayed 
in Fig. 4 for the NL3 parameter set. 
This is mostly responsible for the larger ground state
deformation for $^{150}Sm$.
The somewhat sharper dissolution of the deformation with the
HS parameter set is again traced back to the self-consistent occupancy
evolution with temperature leading to the spherical configuration.
Comparison of Fig. 12 with
Fig. 4 shows that the spin-orbit splitting is significantly larger 
(typically $\sim 40\%$) in case of HS
parameter set as envisaged earlier.

\section{Conclusions}

In this paper, we have attempted to understand in microscopic detail 
the shape evolutions of deformed nuclei with temperature in the RMF
framework. For the rare-earth nuclei we have studied, namely, $Sm$, $Gd$
and $Dy$ with 86 and 88 neutrons, it is seen that the ground state
deformation increases with addition of neutrons appreciably whereas it is
not that sensitive to proton addition.
Similar inference on the sensitivity  of the shape transition temperature
can be drawn to neutron and proton addition. It has been found that  these
transition temperatures are noticeably larger compared to those obtained in
the schematic model with pairing-plus-quadrupole interaction but agree
reasonably well with those calculated \cite{egi2} 
using a realistic force like the Gogny
force. From microscopic viewpoint the ground state deformation
can be understood in terms of the single-particle level structure  near the
Fermi surface. The sudden drop of the deformation can also be understood
from the temperature-driven fast equalisation of the occupancies of
different $K-$states originating from the same $j-$orbital near the
Fermi surface close to the transition temperature.

We have also investigated the role of the non-linear self-coupling of
the $\sigma$ mesons on the shape evolution of the nuclei. 
In absence of the non-linear term, the spin-orbit splitting is enhanced
which affects the single-particle level
structure near the Fermi surface leading to a sharper shape
transition at a lower temperature.

\newpage
\noindent{\bf Figure Captions:}

\begin{itemize}
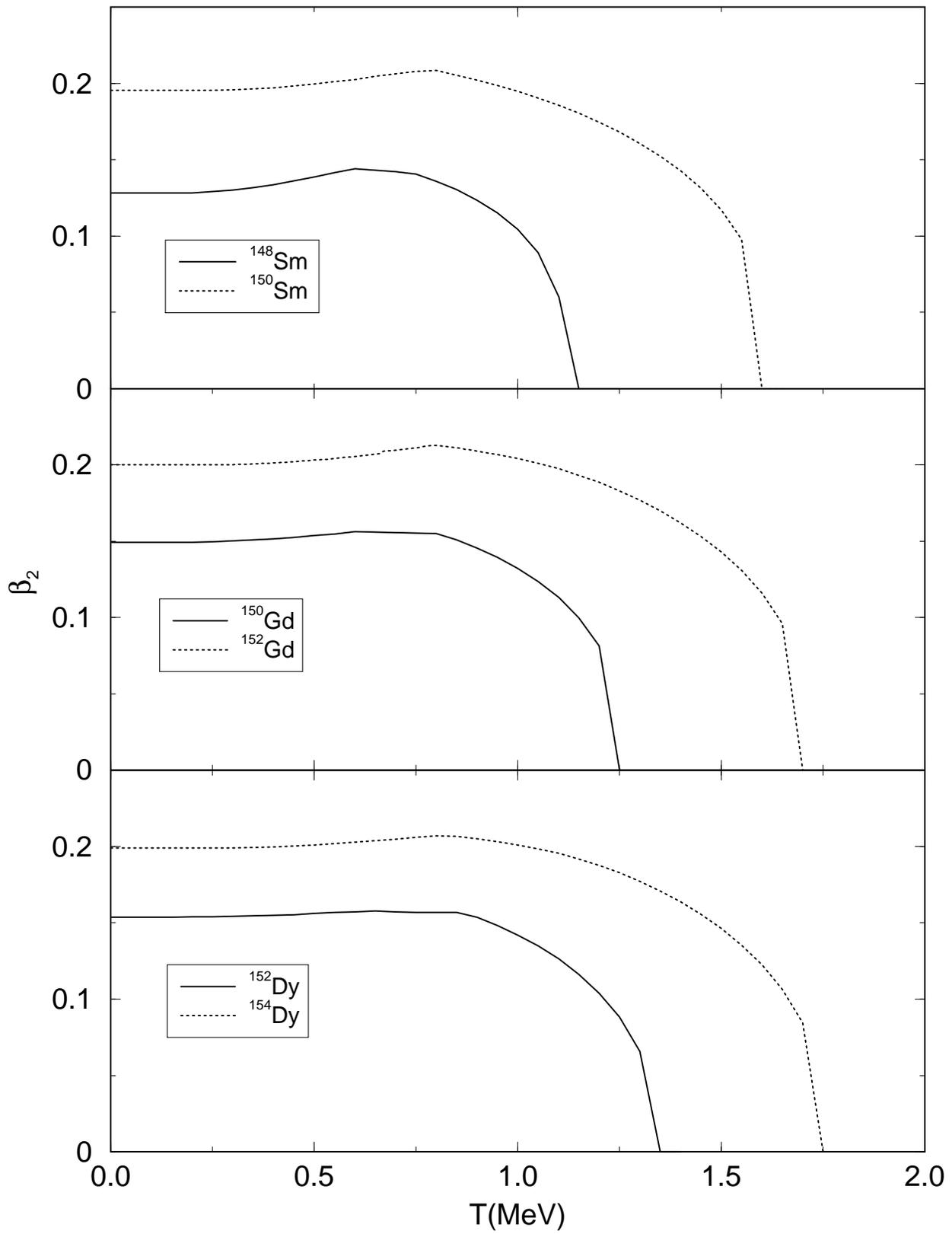
\item[Fig. 1:] The temperature evolution of the deformation ($\beta_2$)
for a few rare-earth nuclei with the NL3 parameter set.
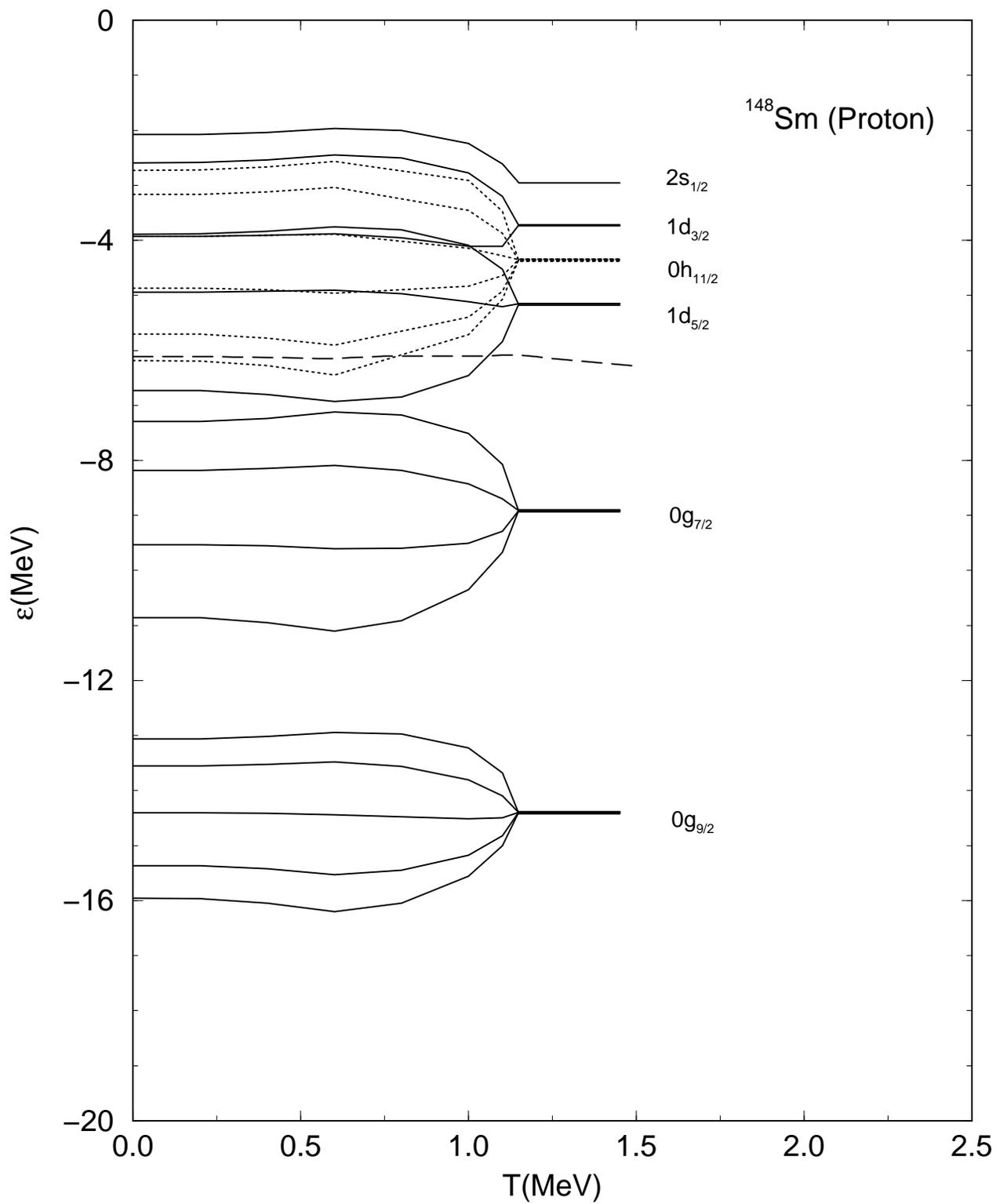
\item[Fig. 2:] The proton single-particle level spectrum near the
Fermi surface for $^{148}Sm$ as a function of temperature with
the NL3 parameter set. The dashed line
corresponds to the Fermi surface.
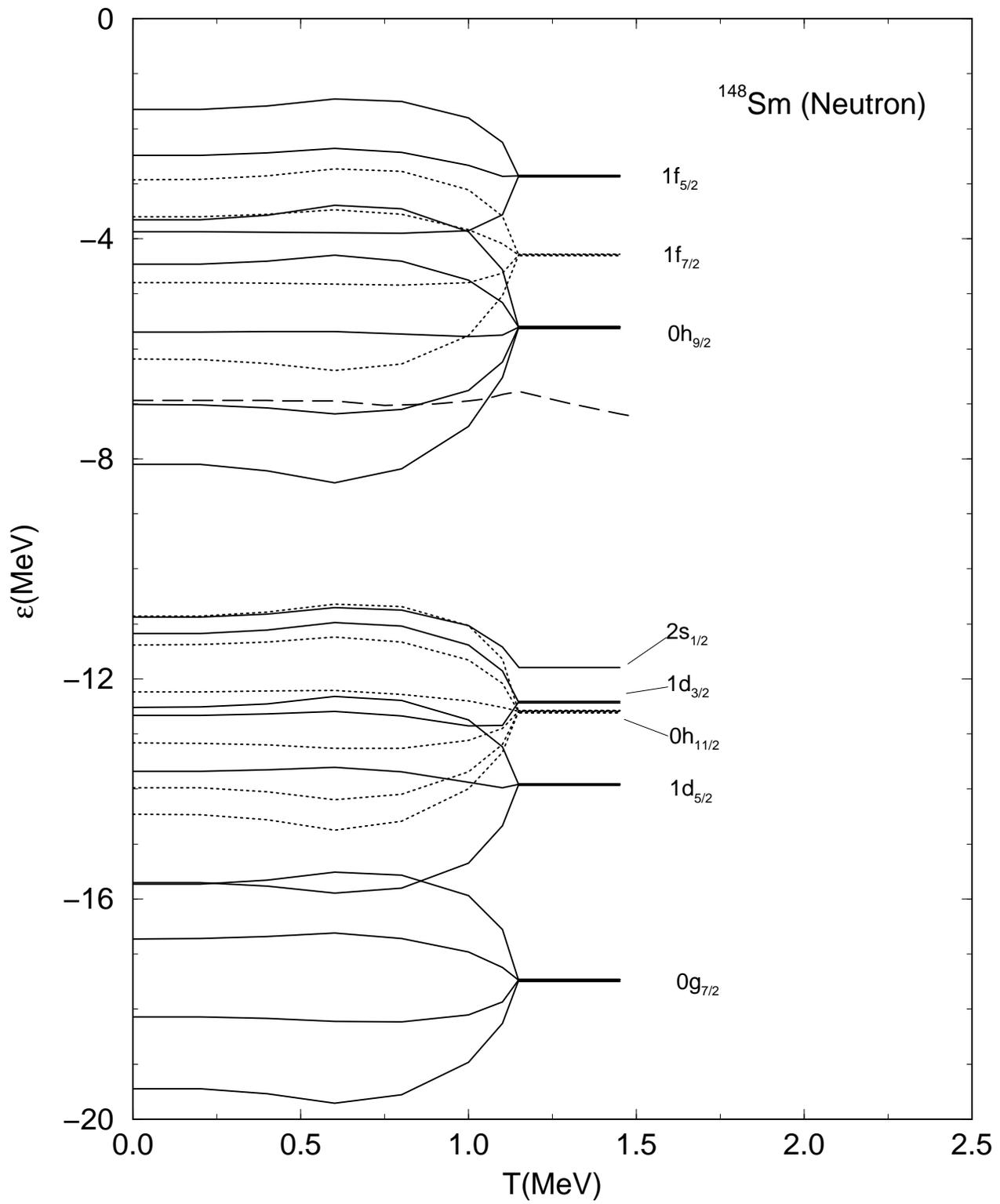
\item[Fig. 3:] The neutron single-particle level spectrum near the
Fermi surface for $^{148}Sm$ as a function of temperature with
the NL3 parameter set. The dashed line
corresponds to the Fermi surface. 
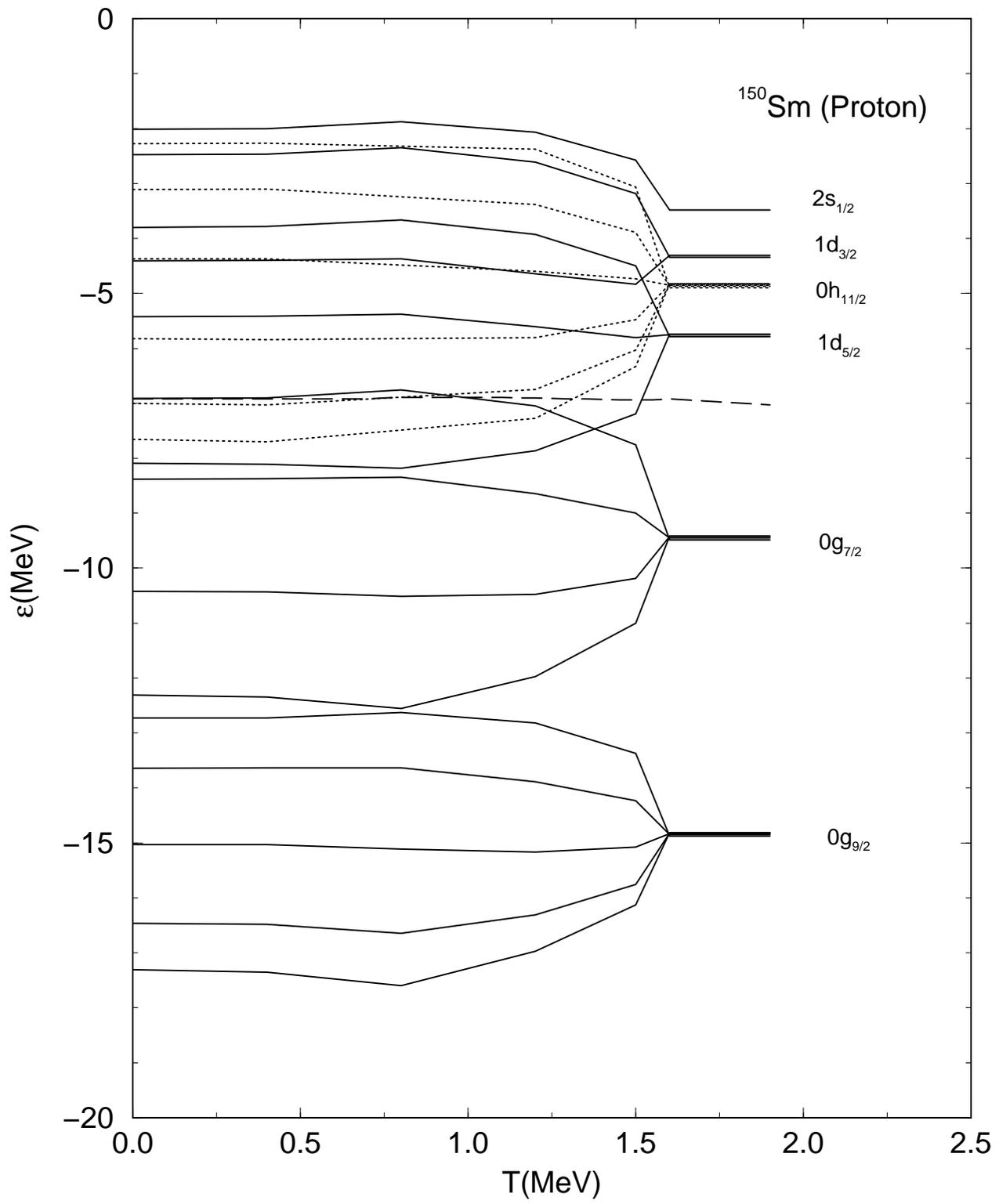
\item[Fig. 4:] Same as Fig. 2 for $^{150}Sm$.
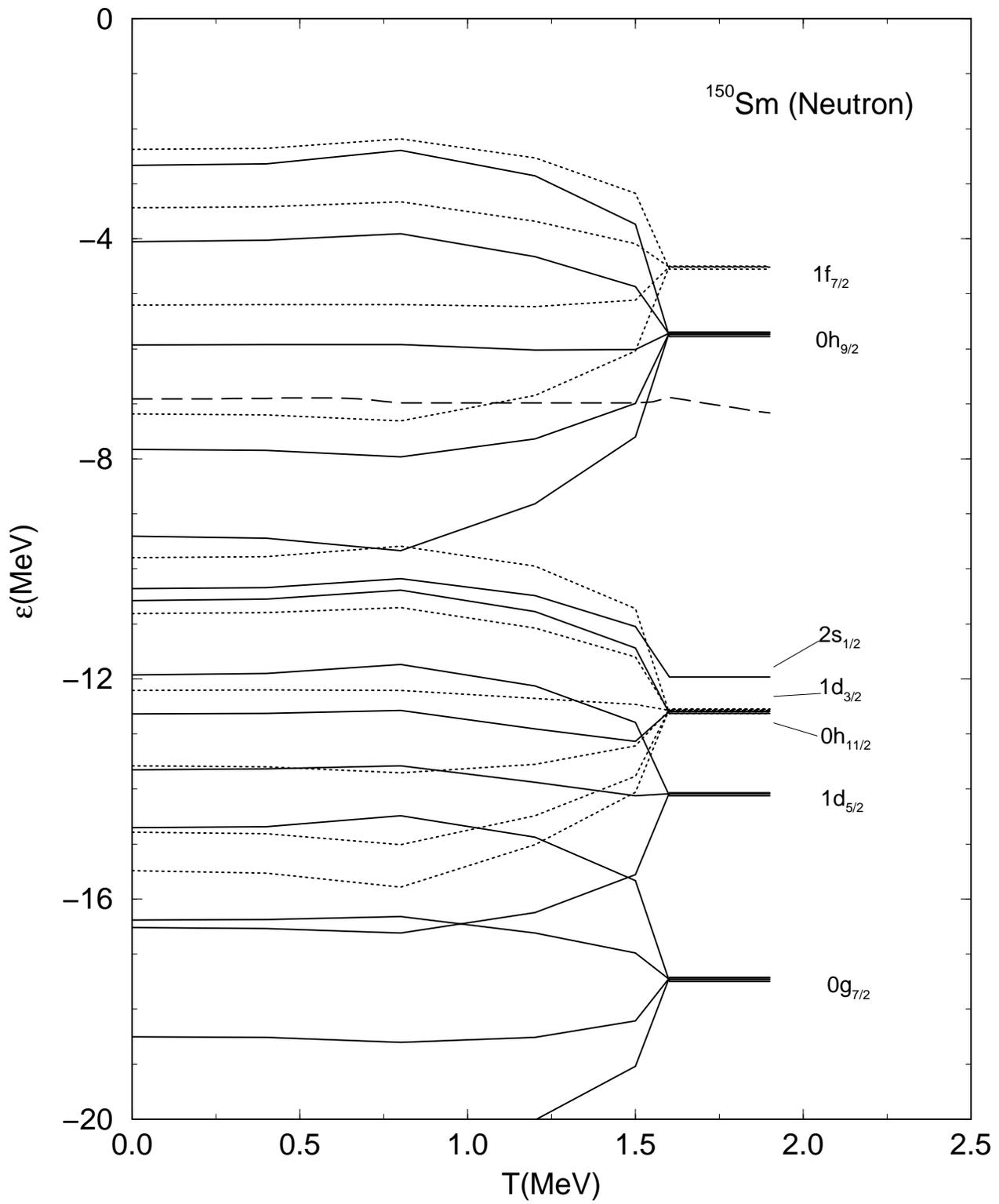
\item[Fig. 5:] Same as Fig. 3 for $^{150}Sm$.
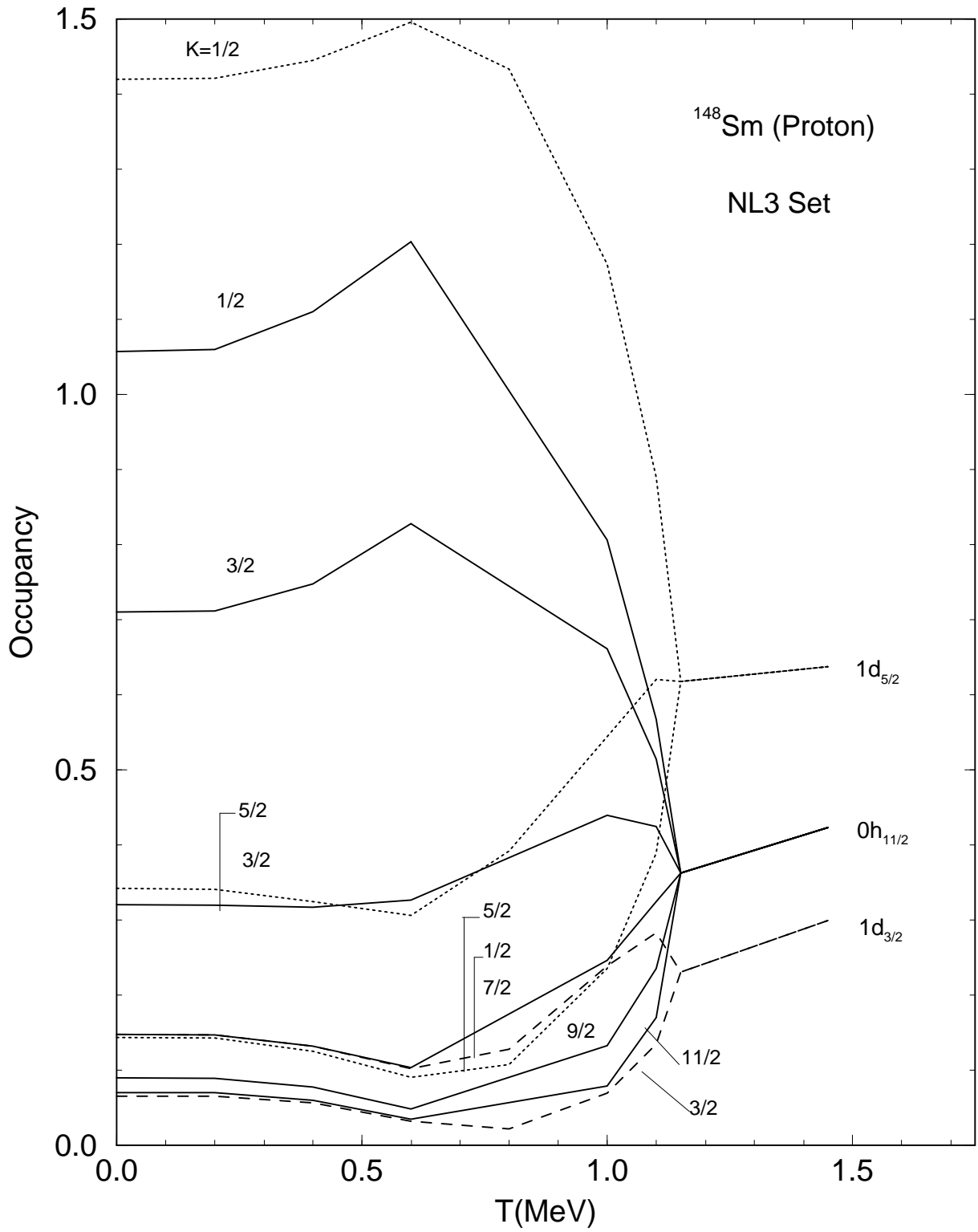
\item[Fig. 6:] The thermal evolution of single-particle occupancy for
protons in  $^{148}Sm$ for a few levels with good projection quantum number
$K$ (as indicated) near the Fermi surface. The degeneracy  of $\pm K$ has
been taken into account in the occupancy.
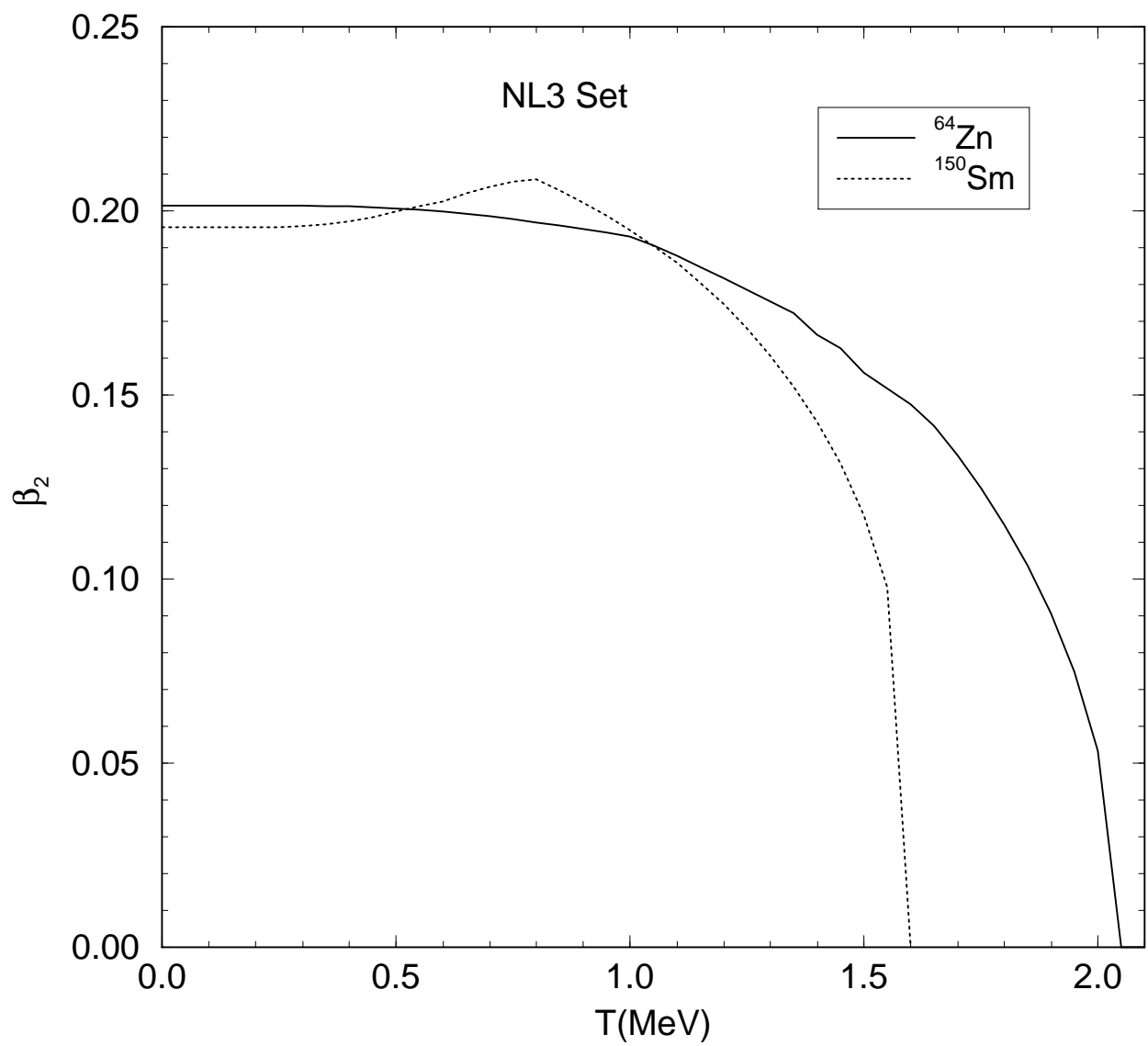
\item[Fig. 7:] Comparison of the thermal evolution of deformation $\beta_2$ for
$^{64}Zn$ (solid line) with $^{150}Sm$ (dotted line).
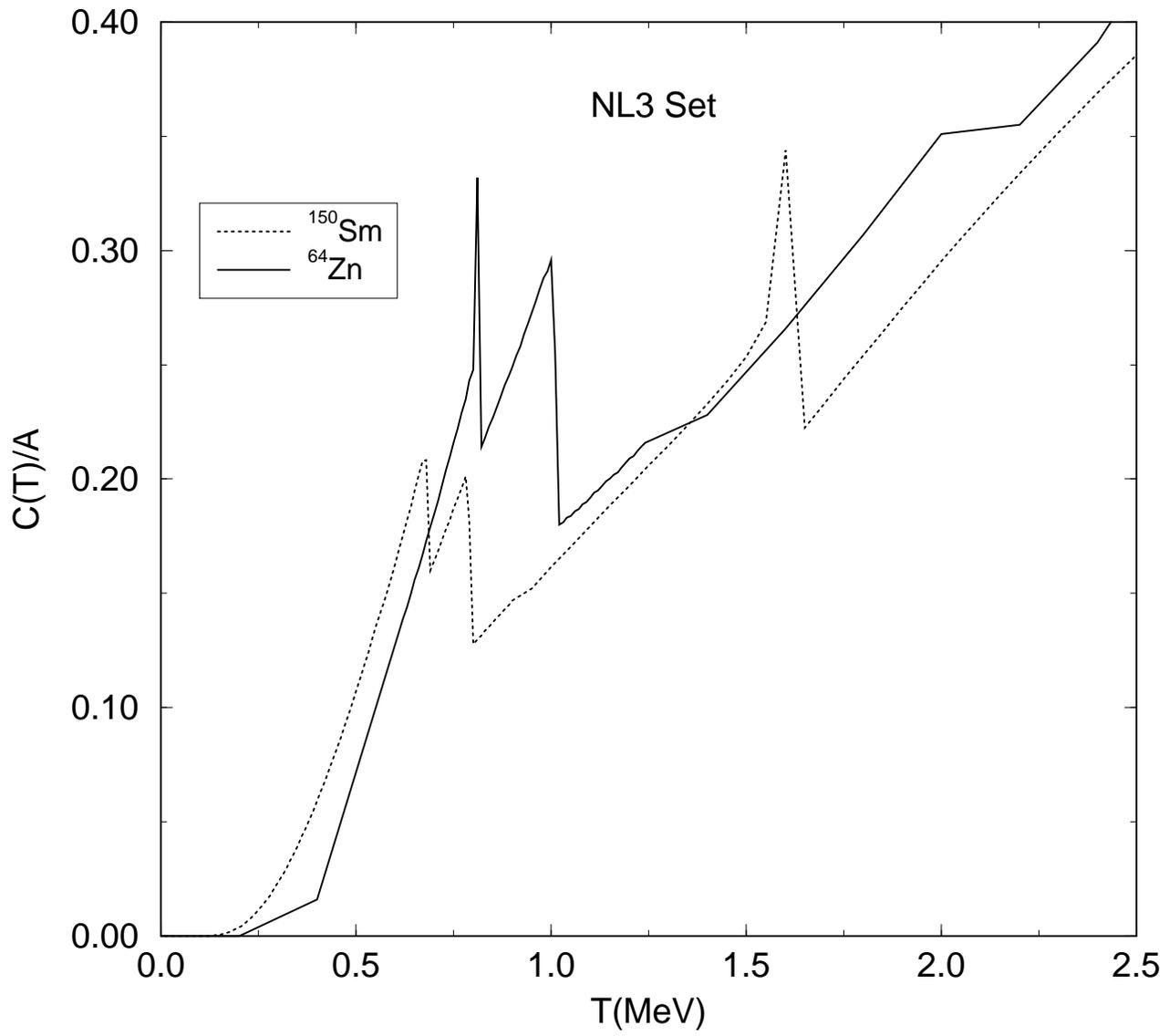
\item[Fig. 8:] The specific heat per particle as a function of
temperature for the systems $^{150}Sm$ and $^{64}Zn$.
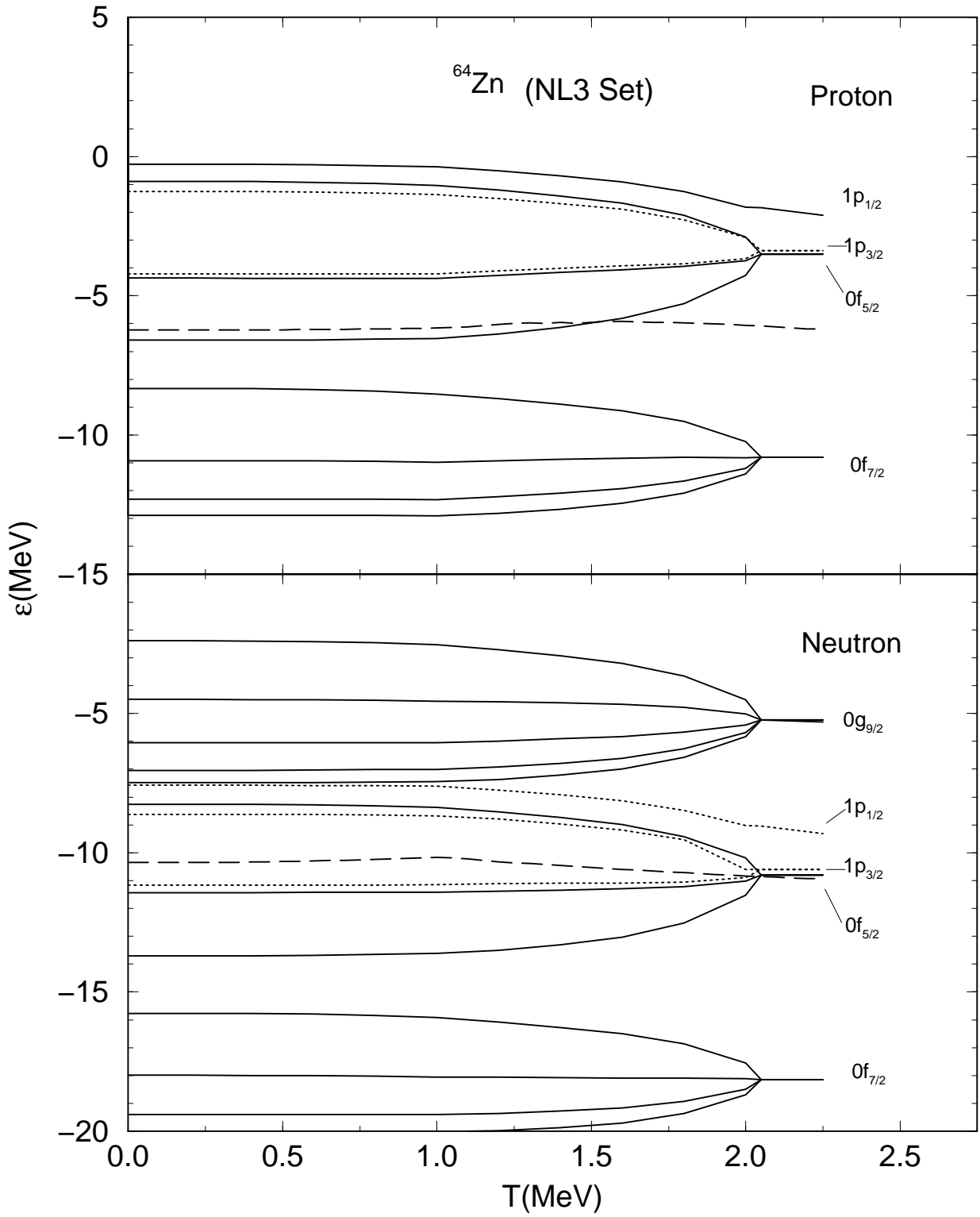
\item[Fig. 9:]  The proton (top panel) and neutron (bottom panel)
single-particle spectra for $^{64}Zn$ near the Fermi surface as a function
of temperature.
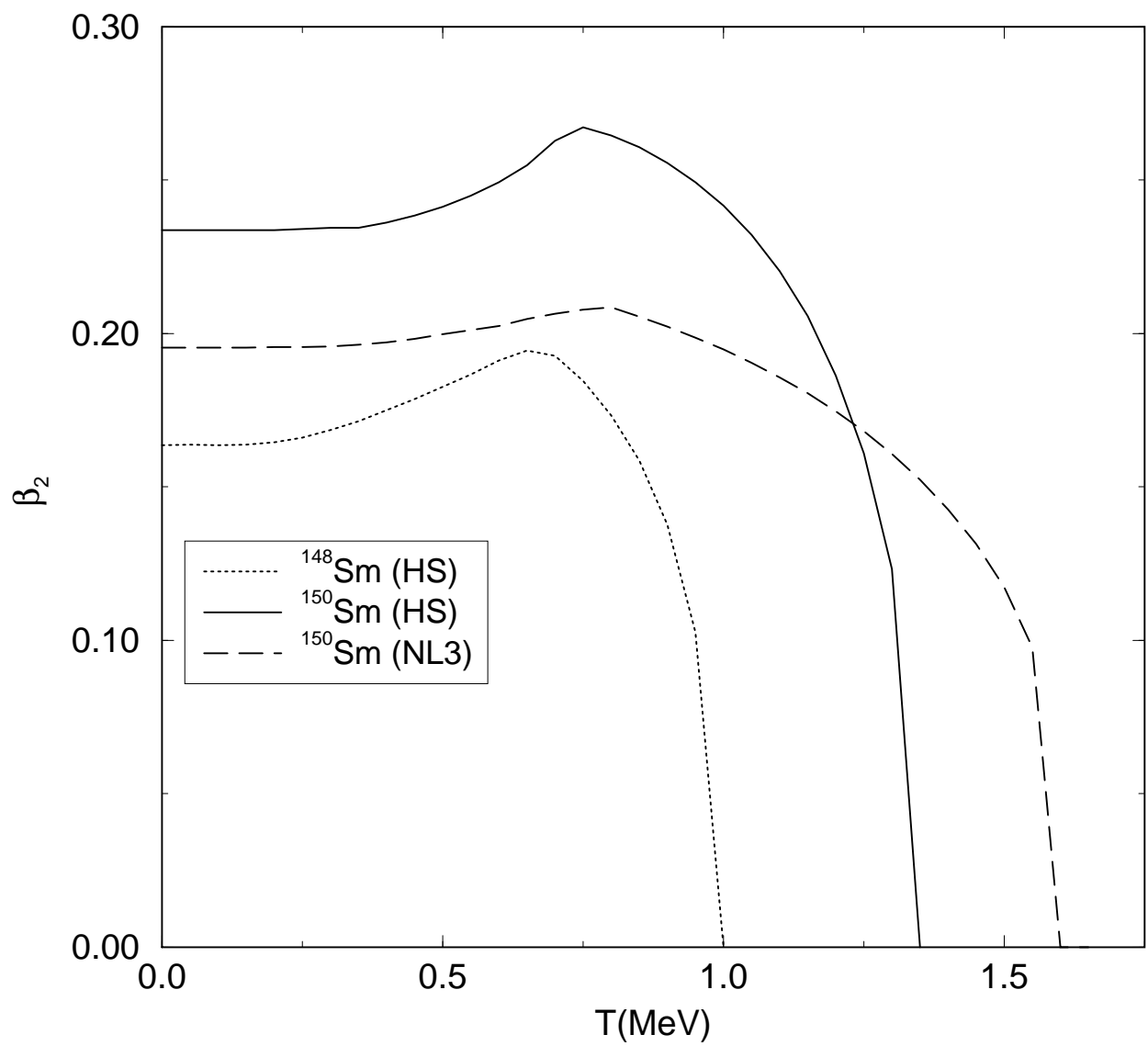
\item[Fig. 10:] The thermal evolution of deformation for $^{148}Sm$
and $^{150}Sm$ with the HS parameter set. For the sake of comparison, the
same for $^{150}Sm$ with the NL3 parameter set is also shown.
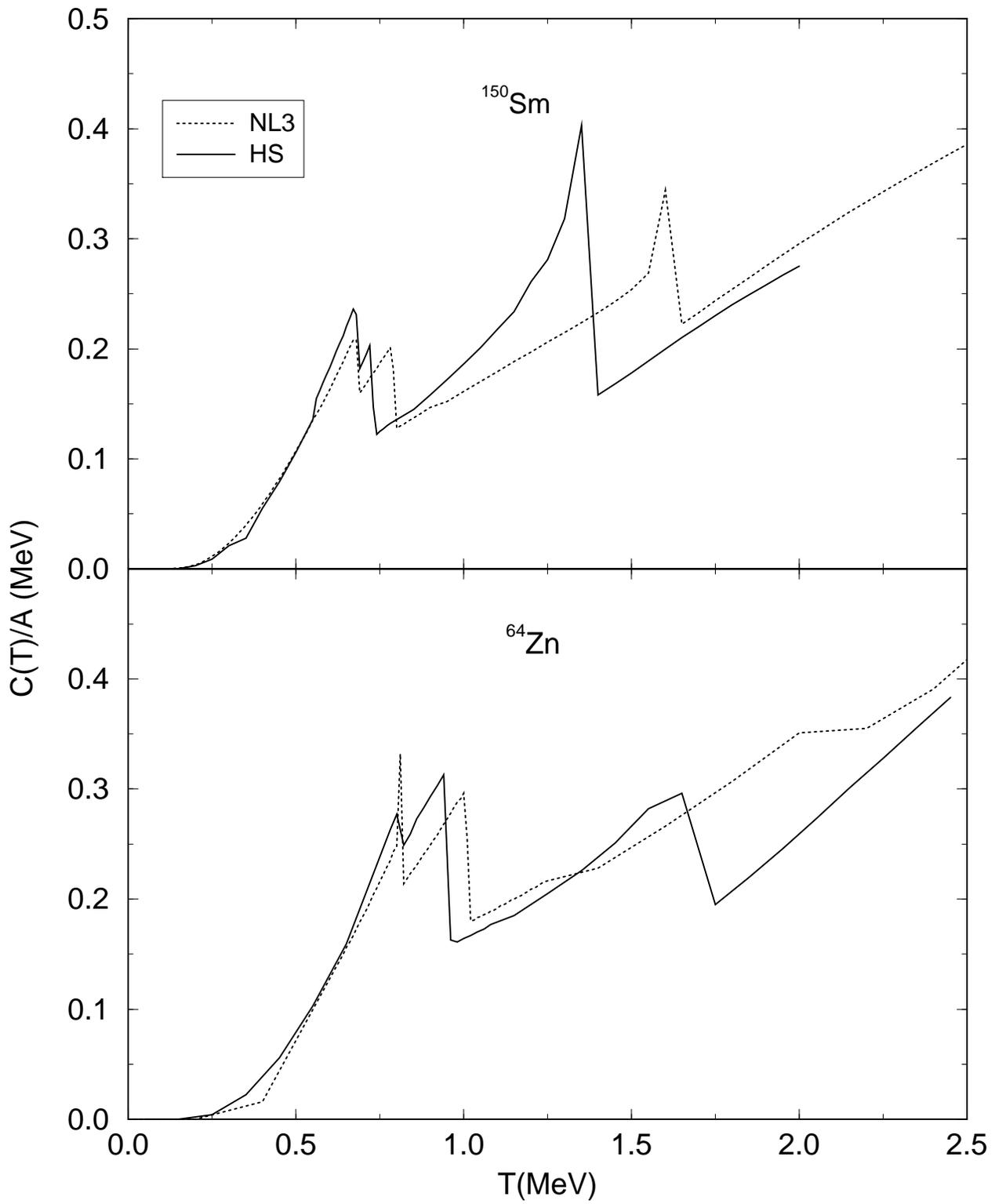
\item[Fig. 11:] The specific heat per particle as a function of
temperature for $^{150}Sm$ with the HS (solid line) and the NL3 (dotted line)
parameter sets (top panel). In the bottom panel, the same is shown
for $^{64}Zn$. 
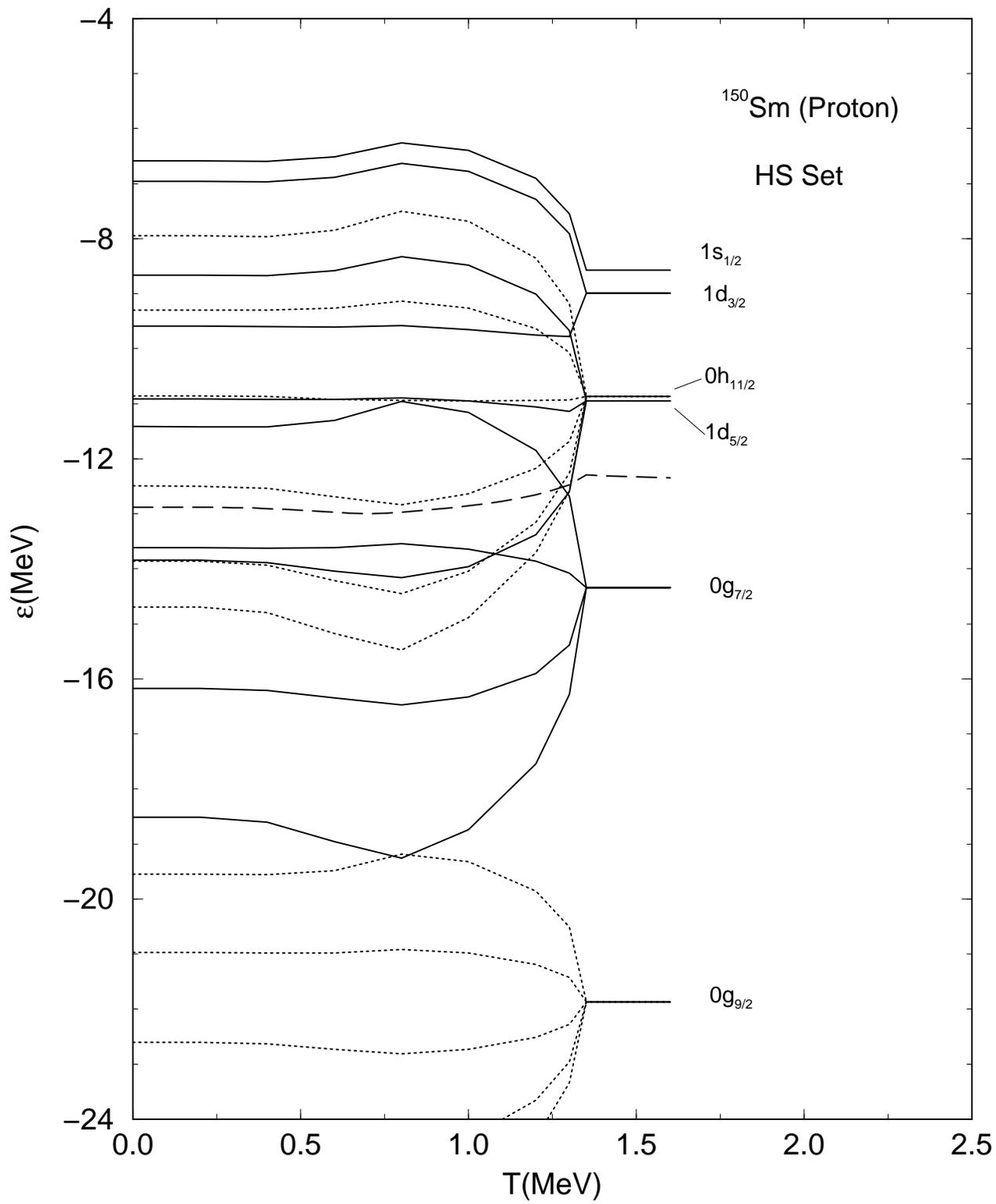
\item[Fig. 12:] The proton single-particle level spectrum near the
Fermi surface with the HS parameter set for $^{150}Sm$ as a function of
temperature.
\end{itemize}

\newpage
\begin{thebibliography}{99}
\bibitem{sno} K. A. Snover, Annu. Rev.  Nucl.  Part. Sci.  {\bf 36}, 545  (1986).
\bibitem{gaa} J. J. Gaardhoje, Annu. Rev. Nucl. Part. Sci. {\bf 42}, 483  (1992).
\bibitem{lev} S. Levit and Y. Alhassid, Nucl. Phys. {\bf A413}, 439  (1984).
\bibitem{alh} Y. Alhassid, J. Zingman and S. Levit, Nucl. Phys. {\bf A469}, 205 (1987).
\bibitem{bra} M. Brack and P. Quentin, Phys. Scr. {\bf A10}, 163 (1974).
\bibitem{que} P. Quentin and H. Flocard, Annu. Rev. Nucl. Part. Sci.
{\bf 28}, 523 (1978).
\bibitem{goo1} A. L. Goodman, Phys. Rev. {\bf C34}, 1942 (1986).
\bibitem{goo2} A. L. Goodman, Phys. Rev. {\bf C38}, 977 (1988).
\bibitem{egi1} J. L. Egido, P. Ring and H. J. Mang, Nucl. Phys. {\bf A451}, 77
(1986).
\bibitem{agr1} B. K. Agrawal, Tapas Sil, J. N. De and S. K. Samaddar, 
     Phys. Rev. {\bf C62}, 044307 (2000).
\bibitem{agr2} B. K. Agrawal, Tapas Sil, S. K. Samaddar and J. N. De, 
     Phys. Rev. {\bf C} (in press).
\bibitem {mil1} H. G. Miller, R. M. Quick, G. Bozzolo and J. P. Vary,
Phys. Lett. {\bf B168}, 13 (1986).
\bibitem{mil2} H. G. Miller, R. M. Quick and B. J. Cole, Phys. Rev.
{\bf C39}, 1599 (1989).
\bibitem{bog} J. Boguta and A. R. Bodmer, Nucl. Phys. {\bf A292}, 413 (1977).
\bibitem{gam} Y. K. Gambhir, P. Ring and A. Thimet, Ann. Phys. (N.Y.)
{\bf 198}, 132 (1990).
\bibitem{rin} P. Ring, Y. K. Gambhir and G. A. Lalazissis, Comp. Phys. Comm.
{\bf 105}, 77 (1997).
\bibitem{lal} G. A. Lalazissis, J. K$\ddot o$nig and   P. Ring, Phys. Rev. {\bf
C55}, 540 (1997).
\bibitem{agr3} B. K. Agrawal, S. K. Samaddar, J.N.De and S. Shlomo, Phys.
Rev. {\bf C58}, 3004 (1998).
\bibitem{bon} P. Bonche, S. Levit and D. Vautherin, Nucl. Phys. {\bf A436},
265 (1985).
\bibitem{egi2} J. L. Egido, L. M. Robledo and V. Martin, Phys. Rev. Lett. 
{\bf 85}, 26  (2000).
\bibitem{ser} B. D. Serot and J. D. Walecka, Adv. Nucl. Phys. {\bf 16},
1 (1986).
\bibitem{hor} C. J. Horowitz and B.D. Serot, Nucl. Phys. {\bf A368}, 503
(1981).
\end{thebibliography}
\end{document}